\documentclass[useAMS,usenatbib,12]{article}
\usepackage[dvips]{graphicx}
\usepackage{color}
\title {\textbf{ BIPOLAR TRANSISTOR TESTER / DIGITAL IC TESTER FOR PHYSICS LAB}}
\author {RAJU BADDI\\ National Center for Radio Astrophysics, TIFR, Ganeshkhind, P.O.Bag 3, PUNE 411007.\\ }
\date{}
\begin{document}

\label{firstpage}

\maketitle

\begin{abstract}
A very simple low cost bipolar transistor tester for physics lab is given. The proposed circuit not only indicates 
the type of transistor(NPN/PNP) but also indicates the terminals(emitter, base and collector) using simple dual 
{\color{red}$\bullet$}/{\color{green}$\bullet$} LEDs.  Color diagrams of testing procedure have been given for 
easy following. This article describes the construction of this apparatus in all detail with schematic circuit 
diagram, circuit layout and constructional illustration. The second part describes a simple circuit to test digital 
ICs in a physics lab. Small scale integration digital ICs like logic gates, flip-flops, registers, counters, decoders, 
multiplexers etc are used in physics lab for various purposes either in a commercially obtained equipment or lab 
made circuits. The non-functionality of a circuit may call for a test of the digital chip. Commercially available 
test equipment is expensive, bulky and some times useless with regard to the concerned test. This article describes 
an inexpensive, portable and useful digital IC test circuit that could be helpful in detecting the actual fault with 
the chip provided the data sheet for the chip is available. In a very convenient and easy way. The article contains 
neat diagrams and illustrations to help the reader build a proper functional digital IC tester.    
\end{abstract}

\section{Bipolar Transistor Tester}
\subsection{Introduction}
Bipolar transistors are frequently used in physics laboratory for a variety of purposes. Some times during experiments 
it becomes necessary to test a transistor for its functioning. Normally this is done using expensive apparatus which 
is microprocessor based and sports a luxurious indication of transistor terminals using alphabets(e, b and c). Here 
a transistor tester is given which uses 3 simple common digital ICs(CD4040, CD4066 and CD4069) but yet is powerful 
enough that it can indicate both the type of the transistor as well as identify its terminals. As such the instrument 
is low cost and easy to build once the details of construction are available(which the author has already done for the 
reader).  Further the testing procedure is also very simple requiring no more than plugging the transistor,  pressing 
a button and observing color indication of {\color{red}$\bullet$}/{\color{green}$\bullet$} LEDs.\\

\subsection{The Bipolar Transistor Tester Circuit}
This tester is primarily meant to test bipolar transistors. It can indicate the type of the transistor as well as identify 
its base, collector and emitter pins. Interestingly the circuit is very simple. The circuit tests conduction in both 
directions, all the possible combinations of three points, meant to plug the transistor taken two at a time. The direction 
of current flow from each point is indicated by a pair of LEDs({\color{red}$\bullet$}/{\color{green}$\bullet$}) associated 
with each point. An NPN(PNP) transistor will produce a 
{\color{red}$\bullet$}{\color{green}$\bullet$}{\color{red}$\bullet$}({\color{green}$\bullet$}{\color{red}$\bullet$}{\color{green}$\bullet$}) 
glow and viceversa according to which test point connects to which terminal of the transistor. Emitter and Collector are 
differentiated by pressing a push-button.  The heart of the circuit is a AC(alternating current) generator  built using 
inverter gates of CD4069. It supplies the AC required to test a pair of points for conduction in both the directions. 
Different combinations are selected by an arrangement of counter (CD4040) and electronic switching(CD4066) which are also 
simultaneously controlled through the AC. \\

\begin{figure}[here]
\begin{center}
\includegraphics[width=130mm,height=115mm,angle=0]{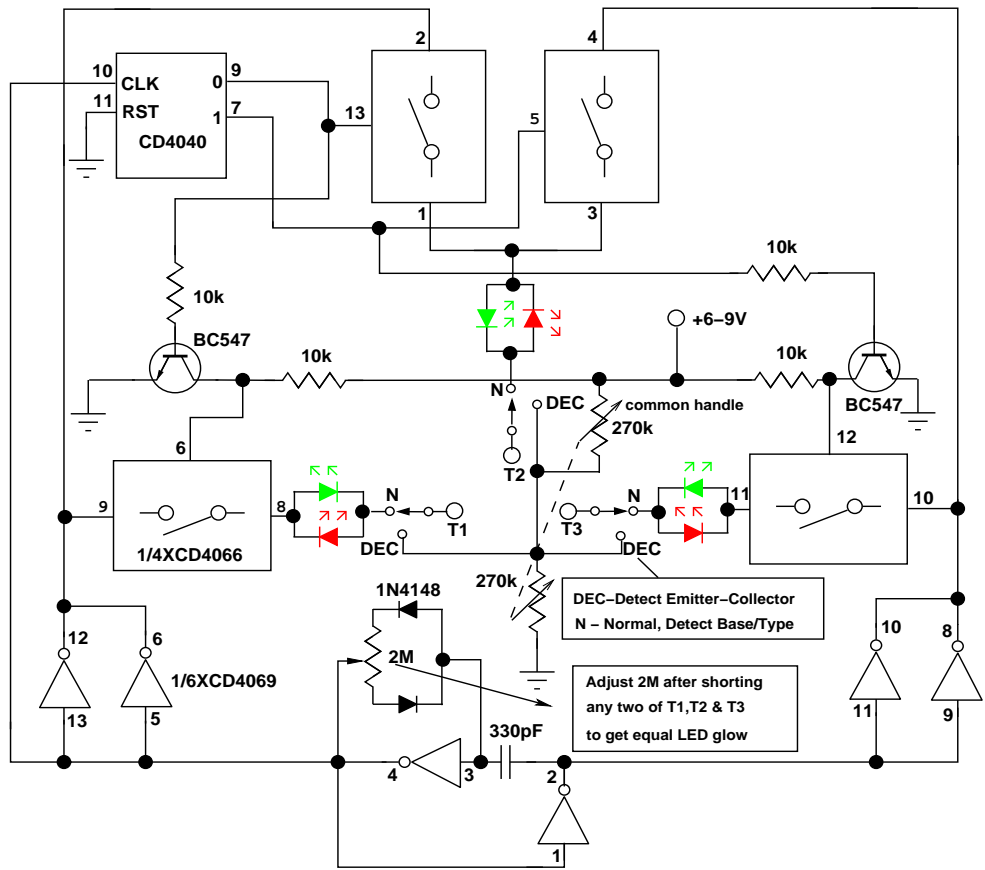}
\caption{Schematic circuit diagram of Bipolar Transistor Tester. This tester uses 
3 simple CMOS chips to make a powerful tester that can not only tell about the type 
of transistor but also about its terminals.}
\end{center}
\end{figure}
\begin{figure}[here]
\begin{center}
\includegraphics[width=75mm,height=80mm,angle=0]{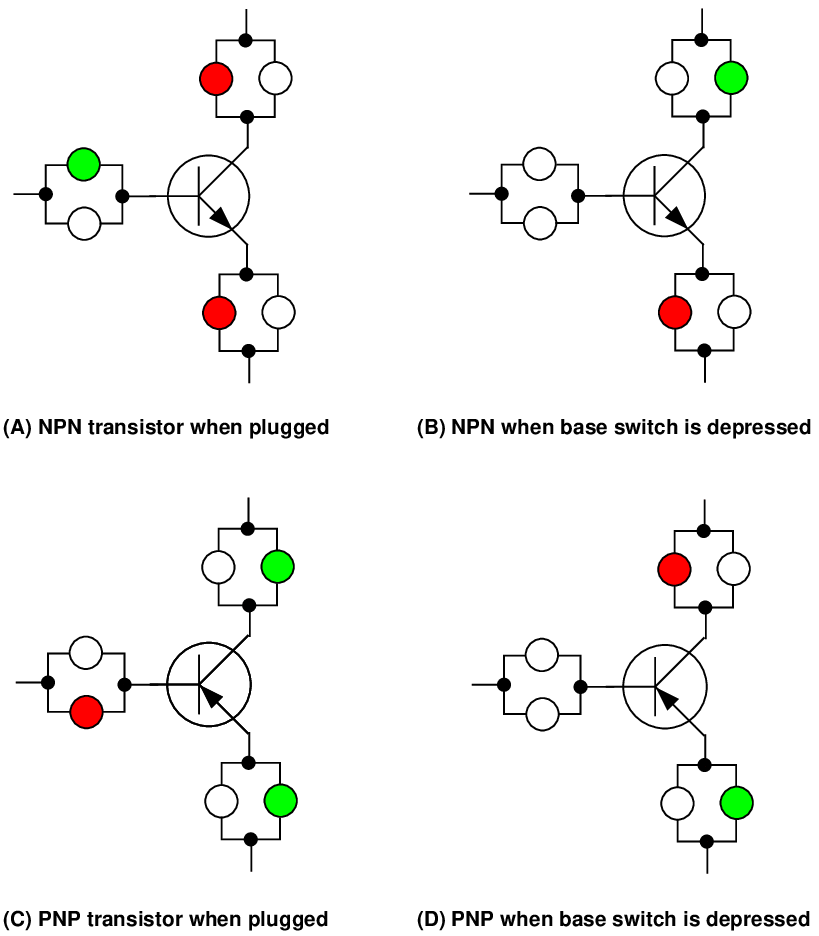}
\includegraphics[width=75mm,height=80mm,angle=0]{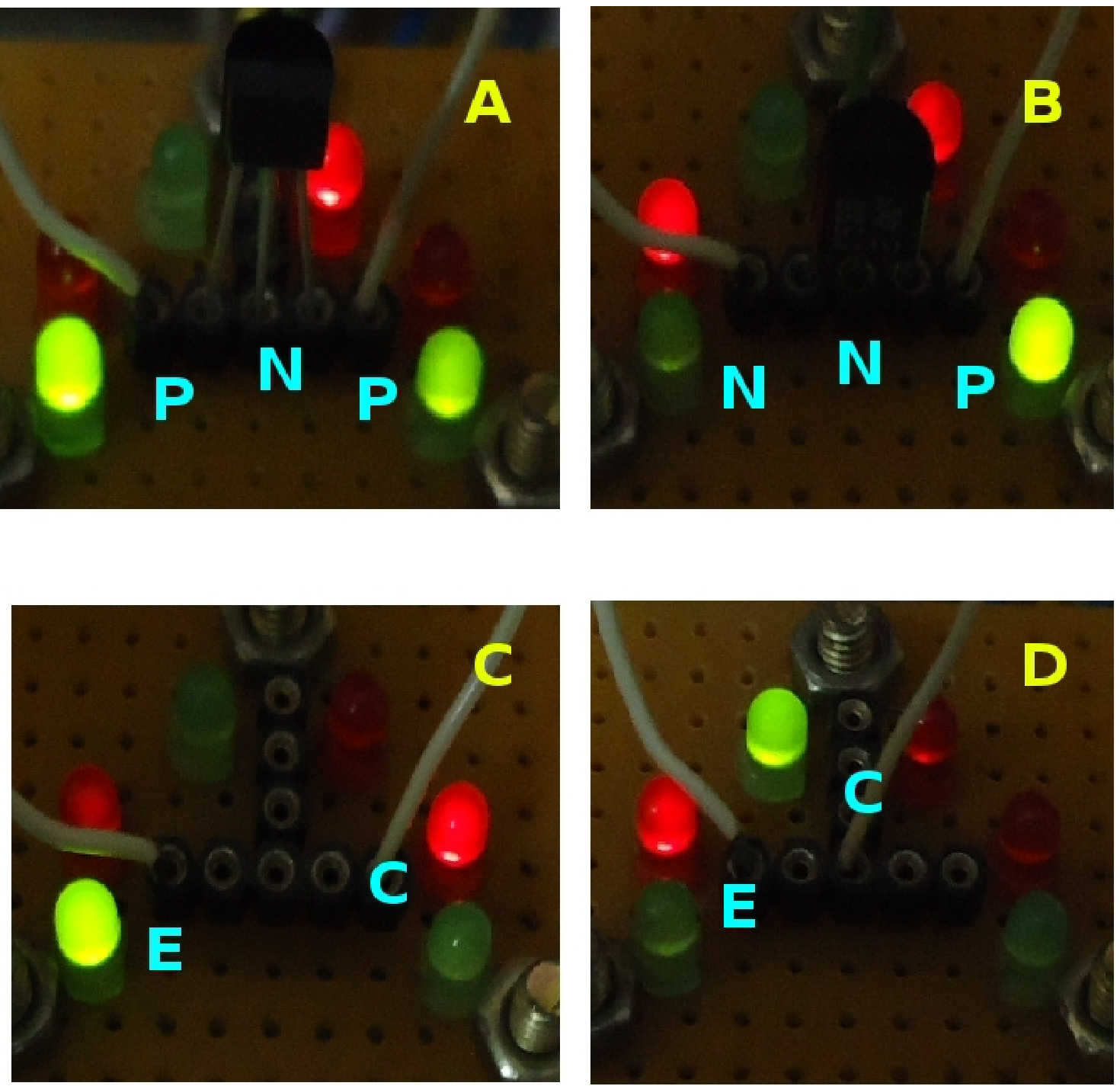}
\caption{Left: Color diagram of LED glow to test the transistor. Right: Photos of PNP/NPN transistor under test. The top 
photo(A-PNP/B-NPN) is for detection of type and base of the transistor, while the bottom(C-PNP/D-NPN) is to 
find emitter and collector. The 2 $\times$ 270k$\Omega$ resistors(potentiometer) can be used to adust the contrast 
during E,C indication.}
\end{center}
\end{figure}

	A pair of LEDs connects to each test point through which current can flow in both the directions. Each LED 
corresponds to a particular direction.  In this manner the two junctions of the transistor are revealed. The LEDs 
are arranged such that they indicate the type of semiconductor across the PN-junction. The counter is clocked by the 
AC generator. So a continuous cycling of combination of pairs occurs. This makes the LEDs glow continuously for easy 
observation revealing the direction of current flow between different test points.  So if the red LED glows connected 
to certain point it means that the N-type of the junction is connected to that test point and viceversa. This reveals 
the type of the transistor NPN({\color{red}$\bullet$}{\color{green}$\bullet$}{\color{red}$\bullet$}) or 
PNP({\color{green}$\bullet$}{\color{red}$\bullet$}{\color{green}$\bullet$}).  From this observation one can easily 
detect the base. To differentiate the collector and emitter use is made of the property that the gain of the transistor 
is polarised i.e for a NPN transistor the emitter should act as emitter of electrons and collector should act a collector 
of electrons. However if one decides to use the collector as emitter and the emitter as the collector then the gain would 
be drastically reduced.   When a normal transistor is plugged into the socket the first step is to note the bright LEDs
(refer Figure 2) these reveal the side of the junction corresponding to each terminal. From this observation one can easily 
deduce the base and type of the transistor.  After this the next step to detect emitter/collector is to disconnect the base 
terminal from the driving circuitry and connect it to the potential divider formed by 2 $\times$ 270K resistors. This makes the 
base terminal to be at the potential approximately half of the supply voltage. Essentially the transistor(PNP or NPN) 
is in turned-on condition. Under this condition if one uses emitter as emitter with right polarity(-ve for NPN/+ve for PNP) 
then the switch(Emitter-Collector) has lower resistance.  However if one uses emitter as collector and vice versa(+ve for 
NPN/-ve for PNP) the switch has higher resistance i.e no LEDs glow or glow very dim. So one sees that for a NPN transistor 
red LED for emitter and green LED for collector glow. While for PNP green LED for emitter and red LED for collector glow 
brightly under base switched condition. The color diagram and photos of actual setup are given in Figure 2. 
Figure 4 shows the circuit layout and construction details.

\subsection{Summary}
This transistor tester can test bipolar transistors and identify their terminals(emitter,base and collector). It is a 
simple and very low cost instrument.\\

{\bf \large{References}} 
\begin{itemize}
\item \ [1] Baddi, Raju, “Transistor Tester Identifies Terminals” EDN, March 17, 2011.
\end{itemize}

\newpage
\section{Digital IC Tester}
\subsection{Introduction}
Digital ICs are complex electronic circuits which operate on logic input/output basis. Basically a set of inputs has 
to produce a set of outputs. In some cases a change in the logic state of an input(s) is expected to produce a 
specific change in the output(s). Commercially available test equipment test digital ICs by generating a set of inputs 
under microcomputer control and similarly check the concerned outputs. A match with the expected result which is 
preprogrammed determines whether the IC chip is good or bad. However it may happen that when the device is faulty the 
user is provided with the sole information that the device is no-good with no information on what is the defect. Also 
this kind of equipment is bulky and expensive. The main problem in testing a digital IC seems to be in setting up a 
logic state configuration for all its inputs and being able to check the outputs. Normally the outputs can be checked 
sequentially one after the other, whereas the various inputs have to be simultaneously applied. The output can be easily 
checked using a logic probe where as the inputs can be logic-0/1 or a transition 1 $\rightarrow$ 0 or 0 $\rightarrow$ 1 
or a continuous train of pulses at a suitable frequency. In most of the cases or atleast for small scale integration 
chips it is sufficient that one input(commonly the clock input) is required to be changed at a time to see a specific 
response. For instance  such is the case for testing gates, flip-flops, counters, shift registers or latches. With this 
approach towards the inputs and outputs of the digital IC it seems to be possible to develop a jig, which receives the 
test chip, that would set up any of the terminals to the required logic state or even power it with any polarity. One 
can see such a jig in Figure 3 towards the lower left corner. As can be seen this jig has simple connections but any 
of the IC terminals can be configured as logic-0/1 and any of the terminals can be connected to +ve or GND of the supply 
voltage(+5V) through a set of 4 jumper post pins. The logic signal can be directly applied to the concerned IC terminal 
and simultaneously the response can be checked on the concerned IC terminal as mentioned earlier. With this approach 
whose technical implementation will be described in the following section one can test several hundred microchips 
provided the internal circuitry is known to the user at block diagram level or atleast a good knowledge of input/output 
terminals exists. The tester described here can also find the input transition voltage for logic levels which is another 
test frequently desired apart from the basic good/no-good health test.       

\begin{figure}[here]
\begin{center}
\includegraphics[width=170mm,height=135mm,angle=0]{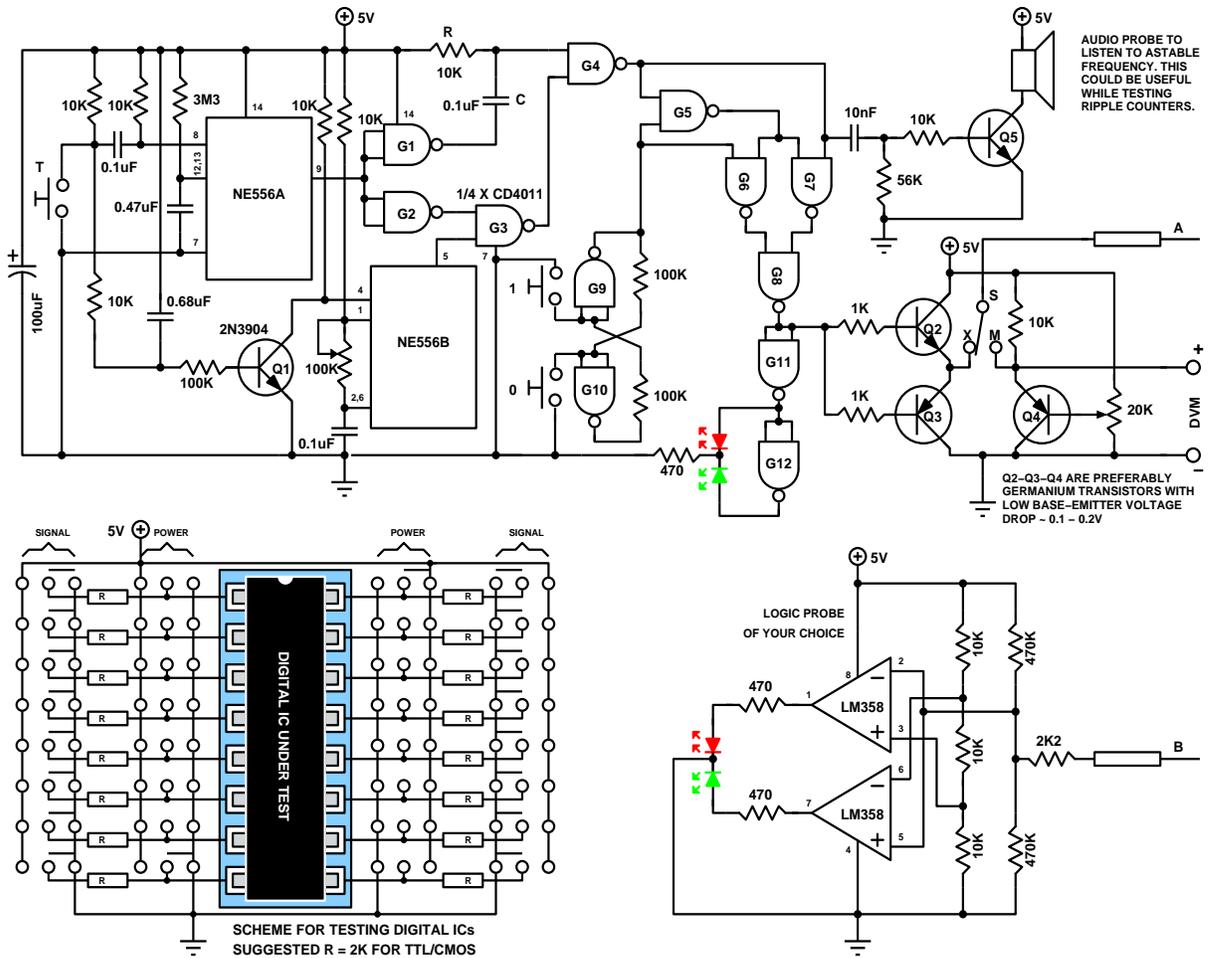}
\caption{Schematic circuit diagram of the Digital IC Tester. The probe A is the pulse probe which is used to inject a 
desired signal to the concerned IC terminal and probe B is the logic probe used to monitor the logic level of the 
outputs of the chip. The jig is the IC receptacle which is a ZIF socket of desired size. The probe A can be switched 
between logic signal(X) and variable voltage follower(M) outputs through the switch S. }
\end{center}
\end{figure}

\subsection{The Digital IC Tester Circuit}
The Digital IC Tester consists of four important sections as can be understood referring Figure 3 .
\begin{enumerate}

\item \ A jig that receives the test chip and can be configured such that any terminal can be set up to a quiescent 
logic state of either 0 or 1 through a resistor(R) so that this state can always be changed when desired by applying 
a logic signal. Also any of the terminals can be either connected to +5V or 0V(GND) supplying the necessary power to 
the chip. These ideal condition logic states can be easily set up through a set of 4 jumper post pins. The jig consists 
of a Zero-Insertion-Force(ZIF) socket and jumper post pins. Figure 9 gives an illustration of the complete IC tester. \\

\item \ The second section is the logic probe towards the lower right of Figure 3(probe B). It is of commonly 
encountered dual operational amplifier(OPAmp) type. Its basic functioning is described in the text. It displays the logic 
state present at its probe tip B on a dual color LED, logic 0/1 $\rightarrow$ {\color{red}$\bullet$}/{\color{green}$\bullet$}. \\

\item \ The third section is the variable voltage generator comprising of a single germanium transistor Q4(as they have 
low base-emitter voltage drop $\sim$0.1-0.2V) configured as voltage follower. This can be used to apply a manually 
controlled variable voltage to the concerned input of the logic device and monitor the output response through the 
logic probe to find the voltage of transition. \\

\item \ The fourth section is the main circuit which produces all the necessary varying logic signals. It comprises of 
NE556 dual timer and three CD4011 NAND chips. The details of the functioning of this section has been described in the 
text. Essentially it has one output(G8) buffered by the voltage followers Q2 and Q3 which can be used to inject logic 
0/1, transition 1 $\rightarrow$ 0, transition 0 $\rightarrow$ 1 or a train of pulses of desired frequency. This section 
is controlled by three switches(labelled T,0 and 1). By pressing 0 or 1 the quiescent state of the output(G8) can be set 
to logic 0 or 1 respectively. By pressing T for a short time one can issue a short single pulse of opposite polarity to 
the quiescent state. Upon prolonged depression, after a short while($\sim$2s) the output starts producing a train of 
pulses whose frequency can be set through a variable resistor. This section forms the pulse injector probe marked A in 
the diagram. The state of the pulse generator probe is indicated by another dual color LED, logic 0/1 
$\rightarrow$ {\color{red}$\bullet$}/{\color{green}$\bullet$}.\\

\end{enumerate}

We first take up the logic probe whose schematic circuit diagram is shown towards the right lower corner of Figure 3. 
It is based upon dual voltage comparator operation amplifiers. Three 10k$\Omega$ resistors in series form a voltage divider 
network whose voltages are used to set upper and lower limits of threshold on the logic-0 and logic-1 voltages that the 
probe would encounter at its input B. The logic-0 voltage should be below 1/3$^{rd}$ of the supply voltage and the logic-1 
voltage should be above 2/3$^{rd}$ of the supply voltage. The probe input B itself is maintained at a voltage of 1/2 of 
the supply through the dual 470k$\Omega$ voltage divider network. The remaining resistors in the circuit are for current 
limiting purpose. When the probe tip B comes in contact with a terminal we assume that the voltage of the terminal is copied 
to the pins 2 and 5 of the OPAmps. Both the comparators now compare this voltage with the biasing voltages from the triple 
10k$\Omega$ network. The lower OPAmp has its inverting terminal biased to 2/3$^{rd}$ of the supply voltage. Its output drives 
the green LED through its anode. For the lower amplifier's output to go high we require the voltage at its non-inverting 
input to be higher than the voltage at its inverting input. So the green LED turns on when the voltage at B is higher than 
2/3$^{rd}$ of the supply voltage. With a similar reasoning we see that the upper OPAmp's output goes high when the voltage 
at B goes below 1/3$^{rd}$ of the supply voltage. In the probe suspended state point B is maintained at 1/2 of the supply 
voltage which lies in between the two limits and hence neither of the OPAmps' output can go high, so neither of the LEDs can 
glow.  \\

We now consider the pulse generator circuit which produces all the necessary types of logic signals for testing. 
One part in NE556 is used as a monostable and the other as an astable. Its astable pulses are used to produce indefinite 
pulse train from the pulse probe. While the monostable is used for the purpose of contact debounce and delay in the 
generation of pulse/pulse-train. The pulse generator probe apart from being stable in either logic-1 or logic-0 state can 
also be made to produce a fixed short time pulse which can be either 0-1-0 or 1-0-1. Over all the circuit has 3 push-button 
switches and 1 SPDT switch. Push button switch 'T' is used to trigger pulse generation while the other two(1/0) are used 
to select the quiescent state(logic-1/0) of the pulse generator. NE556A configured as a monostable
triggers a short pulse generator circuit employing NAND-RC components(G1, R \& C and buffered by G4). The output of the 
monostable is also used to mask(G2,3) the output of NE556B which is configured as astable and provides the indefinite pulse 
train for the probe. When T is depressed for a short time NE556A fires producing the monostable output for $\sim$2 seconds. The 
relatively very short pulse from G1-R-C reaches the pulse probe A through the XOR gate formed by G5-8.  However at the same 
time for about 2s the output of the astable is masked from reaching the XOR. It should be noted that to avoid any spurious 
pulse reaching the probe A NE556B is ideally kept deactivated by applying a low voltage to its reset pin(pin-4) through the 
transistor Q1 whose biasing is further guarded by a 0.68$\mu$F capacitor.  Only upon prolonged depression is NE556B activated 
through the cutoff of Q1. It is only then it starts producing astable oscillations. These reach the probe A only after NE556A 
has completed its monostable period. G9,10 form a simple bi-stable circuit which controls the operation of XOR. G11,12 
together drive the DCL(dual color LED) and take care of the display of the pulse probe state/polarity. The output(eventual) 
of G8 is buffered by a NPN-PNP germanium transistor pair(Q2,Q3) to boost the output current which is essential for 
TTL devices. Germanium transistors have a lower base-emitter voltage drop($\sim$0.1-0.2V) compared to silicon transistors
($\sim$0.6-0.7V) and hence voltage followers employing germanium transistors can produce voltages much below 0.6V which is 
near to the upper limit of logic-0 threshold for TTL devices. The probe A can also be connected to a variable voltage source 
through switch S. The variable voltage source comprises of a germanium PNP transistor whose base is biased by a potentiometer
(20k$\Omega$). The frequency of the NE556B astable can be varied to generate either low frequency(concerns flip-flops or 
small counters) or high frequency(large counters) according to the requirement. This frequency can be easily assessed using 
an audio probe(Q5) as shown in Figure 3 towards the right upper corner. Figures 7-9 show the constructional details of the 
Digital IC Tester and Figure 10 shows a few examples of test set up.  

\subsection{Summary}  

The Digital IC Tester described here can test hundreds of digital ICs conveniently when the internal block diagram or the pin 
connections to the chip are known. This tester is low cost, effective and easy to build/handle. It can provide adequate information 
regarding the defect in the IC chip. For example if one or two gates in a 74LS00 chip were non-functional it can happen that 
the remaining gates are intact/useable. In such a case a conventional tester would fail the chip. The interested user can 
still exploit the chip. A probe type arrangement of the tester can be found in reference [1] of this section. A more simple 
probe type arrangment is also given in Appendix II which uses a single 4011 chip and provides all the basic digital 
functionalities of the circuit in Figure 3. 

\section{APPENDIX I: Circuit layouts and Constructional illustrations}
\begin{figure}[here]
\begin{center}
\includegraphics[width=140mm,height=165mm]{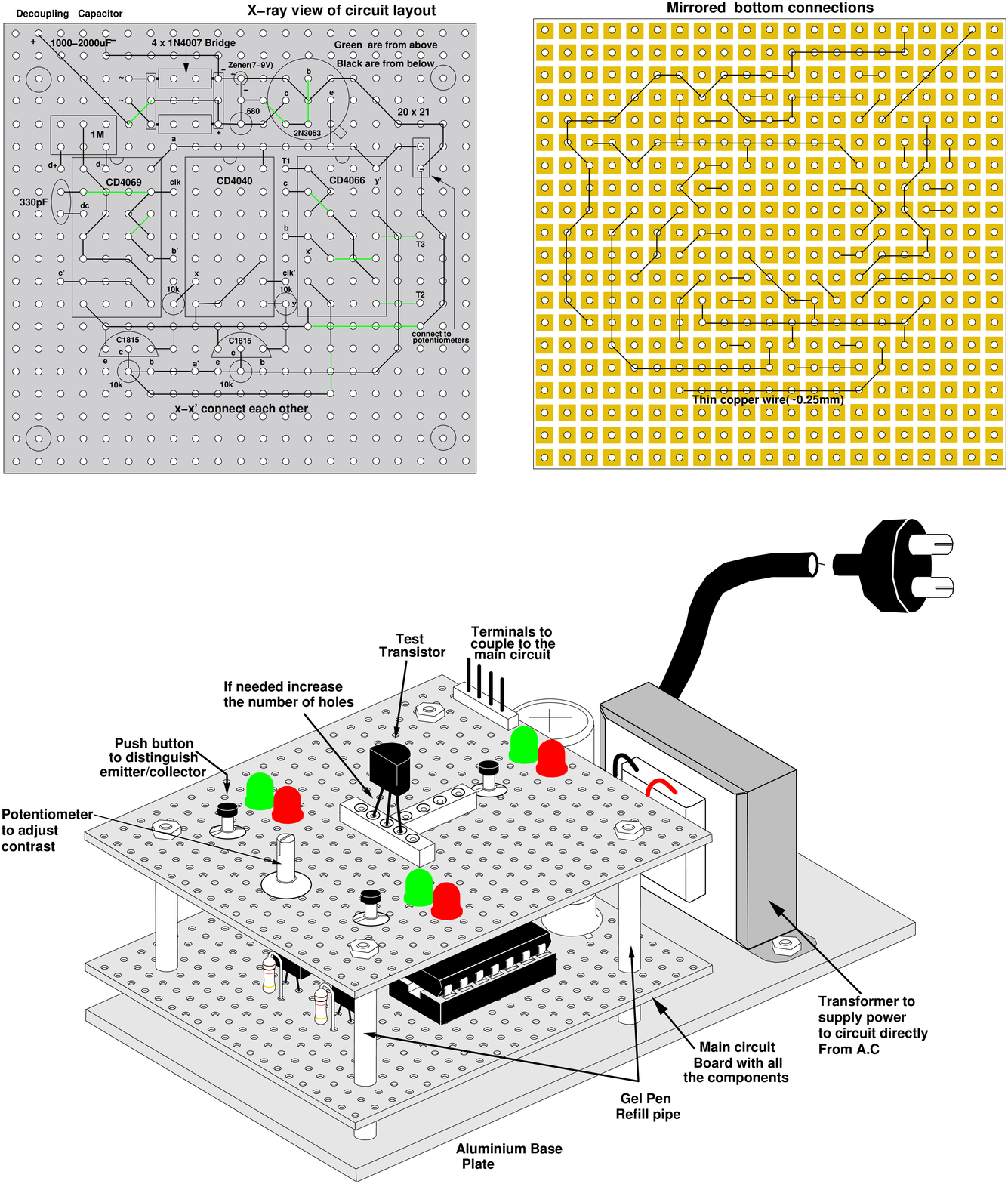}
\caption{Top: Ciruit layout of Bipolar Transistor Tester on a general purpose circuit board. Green connections are 
from top while the black ones are from below $\sim$0.3mm copper wire. Observe the schematic(Figure 1) carefully 
and make any remaining connections. Bottom: Illustration showing constructional details of the Transistor Tester.}
\end{center}
\end{figure}

\section{APPENDIX II: A Handheld Probe type Arrangment of IC Tester}
  Here a handheld probe type arrangement for the Digital IC tester is given which can perform all kinds of digital tests described above 
and is far less complex, employing only one HEF4011BP NAND chip(Figure 5-6). The pulse generator part is constructed out of only 
two NAND gates, whereas the remaining two gates form a well defined voltage window logic probe as given in reference [2] below. The two probes 
behave very similarly to the arrangement in Figure 3. However the maximum frequency of operation is strictly restricted towards 
the lower end. Also one has to be careful while operating this hand held probe. The simplicity comes at a cost of 
certain constraints and ease of operation/arrangement. S is used to toggle between the quiescent states of logic-0/1 and T is 
used to issue a short single pulse or pulse train depending on how its depressed. A video of behaviour of this probe can be 
seen at http://youtu.be/OZ627Rtug\_U in its third part. 

\begin{figure}[here]
\begin{center}
\includegraphics[width=160mm,height=38mm,angle=0]{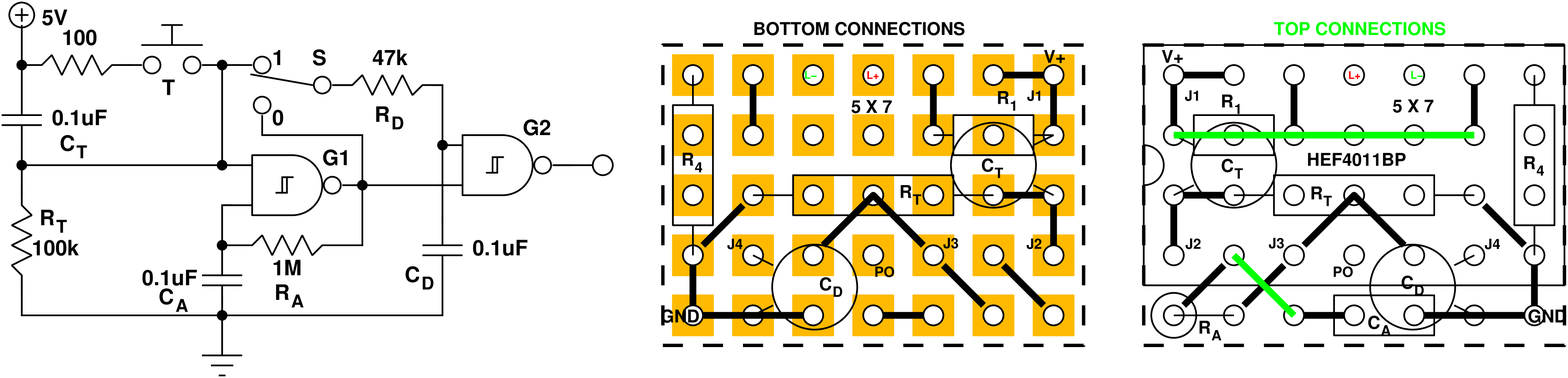}
\caption{The left side shows the schematic circuit diagram of the simple digital IC tester probe. The right side shows the layout
for its construction on a general purpose circuit board. Both X-ray view of component layout and mirrored bottom connections are 
shown. Typical values for dual gate Logic Probe[2], R$_1$=1M$\Omega$,R$_2$=680k$\Omega$,R$_3$=220k$\Omega$ and R$_4$=1M$\Omega$ for 
threshold voltage of V$_T$=2.5V. }
\end{center}
\end{figure}

\begin{figure}[here]
\begin{center}
\includegraphics[width=165mm,height=65mm,angle=0]{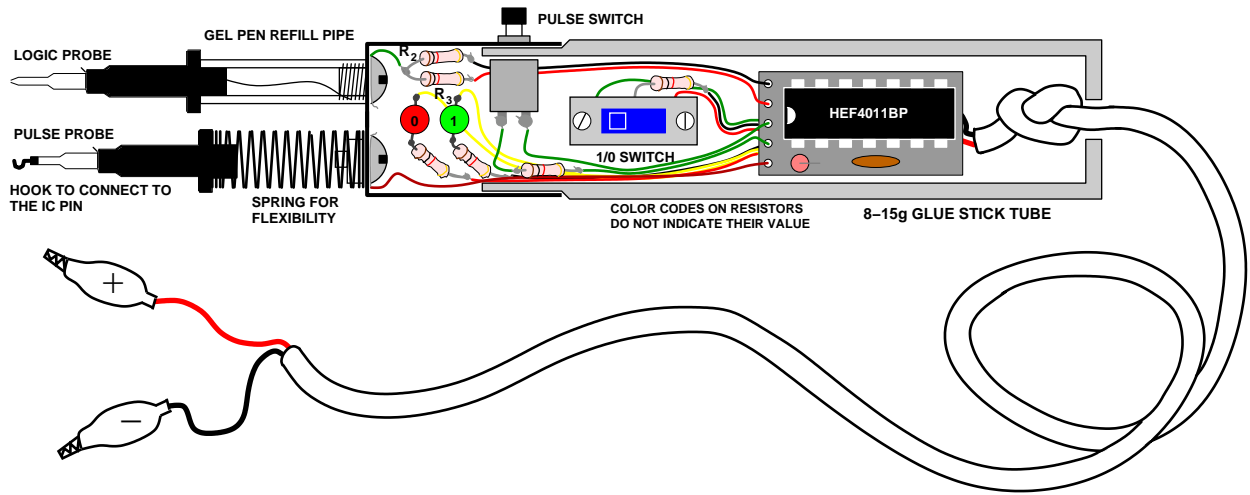}
\caption{Constructional details of the hand held digital IC tester probe in a empty container of glue stick tube(8-15g). The 
pulse switch(T) and the queiscent state switch(S) are at 90$^o$ to each other. R$_2$=680k$\Omega$ and R$_3$=220k$\Omega$ correspond 
to the logic probe[2]. They are in series with the R$_1$ and R$_4$ resistors. In other words their black and red wires connect to 
the pin-12 and pin-8 of HEF4011BP. The LEDs have 470$\Omega$ resistors in series. {\color{red}L-}/{\color{green}L+} are connected to 
GND/+5V respectively.}
\end{center}
\end{figure}
{\bf \large{References}} 
\begin{itemize}
\item \ [1] Baddi, Raju, “Probing system lets you test digital ICs,” EDN, June 21, 2012, pg 48.
\item \ [2] Baddi, Raju, “Single hex-inverter IC makes four test gadgets,” EDN, July 30, 2012, pg 49.
\end{itemize}
\begin{figure}[here]
\begin{center}
\includegraphics[width=230mm,height=115mm,angle=90]{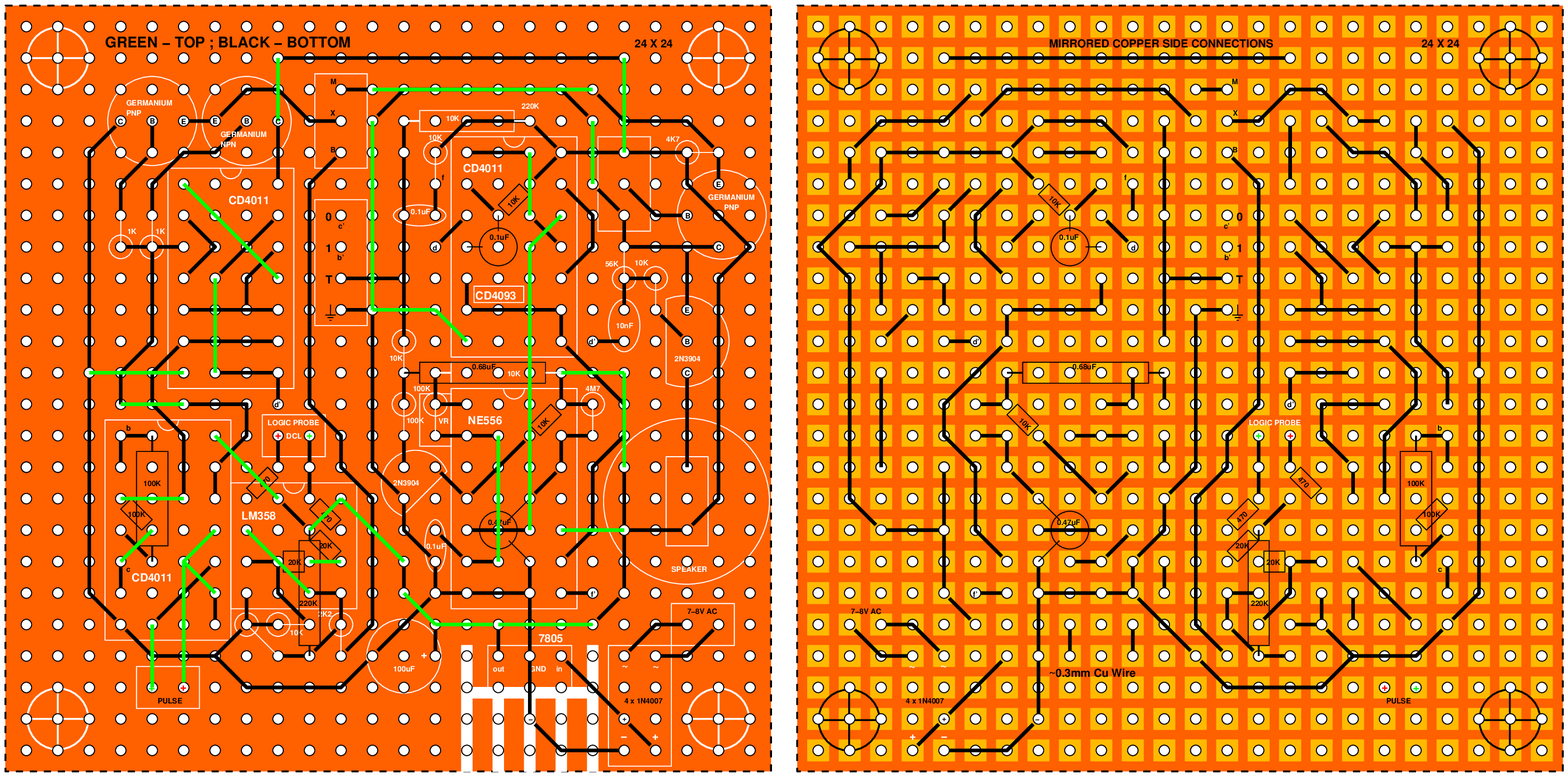}
\caption{Circuit layout of Digital IC Tester lower main board. Use a decoupling capacitor 
of 1000-2200$\mu$F(not shown in the diagram but shown in the illustration) at the output of the 4 $\times$ 1N4007 bridge.}
\end{center}
\end{figure}

\begin{figure}[here]
\begin{center}
\includegraphics[width=230mm,height=143mm,angle=90]{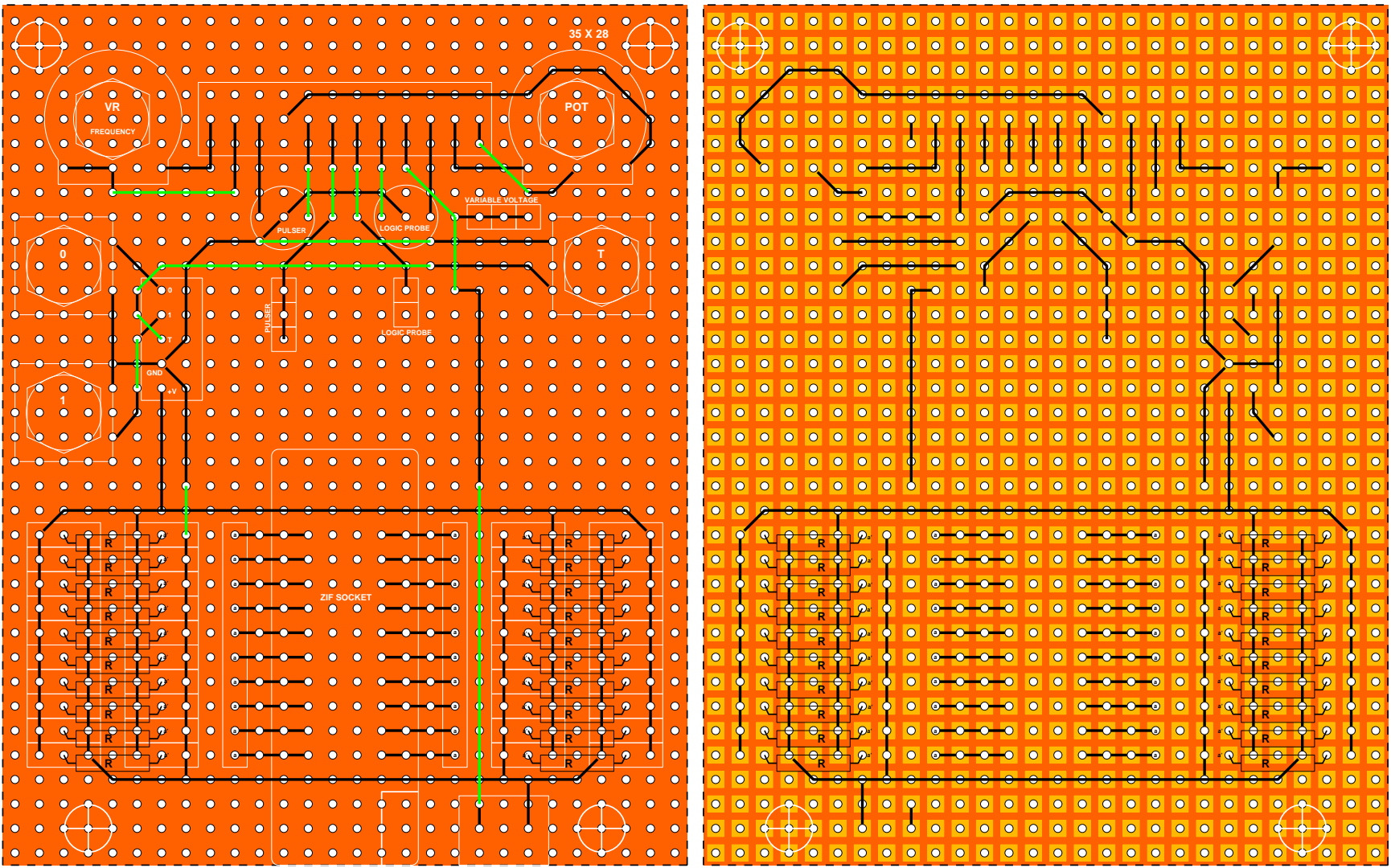}
\caption{Circuit layout of Digital IC Tester upper jig board.}
\end{center}
\end{figure}
 
\begin{figure}[here]
\begin{center}
\includegraphics[width=220mm,height=175mm,angle=90]{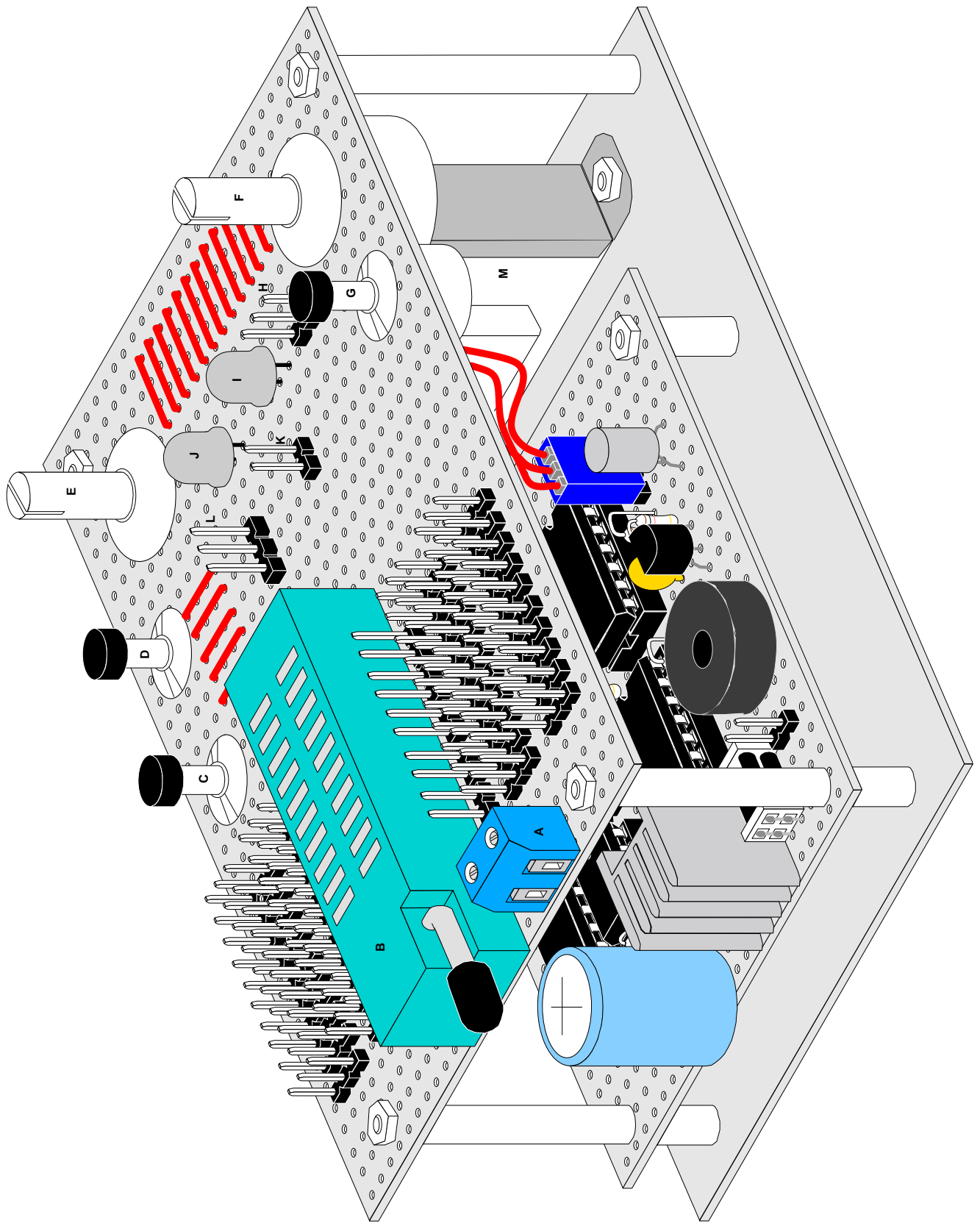}
\caption{Illustration showing the completed Digital IC Tester. A - Voltage measurement 
port, B - ZIF IC insertion socket, C - quiscent logic-1 pulser, D - quiescent logic-0 
pulser, E - Frequency control for NE556B, F - Voltage control for measuring gate's 
transition input voltage threshold, G - Generate toggle pulse, H - PNP voltage follower 
output, I - Logic probe DCL, J - Pulser DCL, K - Logic probe port/tip, L - Pulser probe 
port/tip. M-Transformer for power supply. One set of single post pins is provided on each 
side to establish connection with the IC terminals.}
\end{center}
\end{figure}

\begin{figure}[here]
\begin{center}
\includegraphics[width=165mm,height=230mm,angle=0]{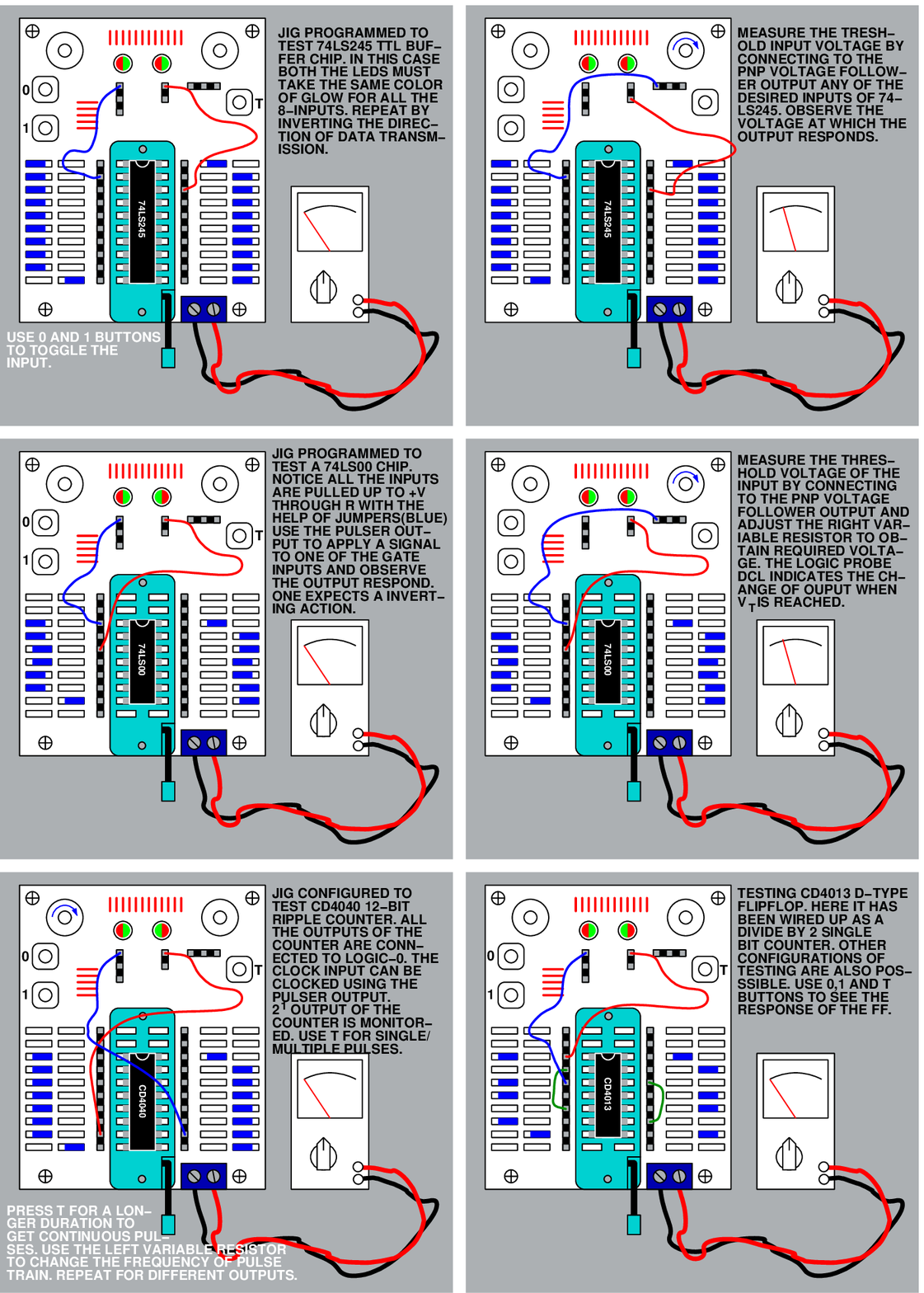}
\caption{Illustration of test setups for some of the ubiquitously encountered simple digital chips.}
\end{center}
\end{figure}

\label{lastpage}

\end{document}